\documentclass[unnumsec,webpdf,contemporary,large]{oup-authoring-template}%

\usepackage{graphicx}
\usepackage{times}
\usepackage{xcolor}
\usepackage{colortbl}
\usepackage{eurosym}
\usepackage{hyperref}
\usepackage{rotating}
\usepackage{multirow}
\usepackage{multicol}
\usepackage{natbib}

\newcommand{\COT}{CO$_2$e}

\newcommand{\tCOTe}{tCO$_2$e}
\newcommand{\tceq}{tCO$_2$e}
\newcommand{\totGoodConf}{258} 
\newcommand{\totGoodSchool}{42} 
\newcommand{\totColdCaseConf}{49} 
\newcommand{\totColdCaseSchool}{13} 
\newcommand{\totMeetings}{362} 
\newcommand{\totGoodMeetings}{300} 
\newcommand{\totColdCaseMeetings}{62} 

\newcommand{\totColdCaseConfEmissions}{4154$\pm$349\,\tceq}
\newcommand{\totColdCaseSchoolEmissions}{365$\pm$50\,\tceq}
\newcommand{\totGoodEmissions}{38,040\,\tCOTe}
\newcommand{\totColdCaseEmisisons}{4,519$\pm$353\,\tceq}
\newcommand{\totEmissions}{42,559$\pm$353\,\tceq} 
\newcommand{\totEmissionsRounded}{42,500\,\tceq} 
\newcommand{\emissionsPerCapConf}{1.1$\pm$0.6\,\tCOTe} %
\newcommand{\emissionsPerCapSchool}{0.7$\pm$0.4\,\tCOTe} %
\newcommand{\emissionsPerCapMeetings}{1.0$\pm$0.6\,\tCOTe}

\begin{document}

\journaltitle{PNAS Nexus}
\DOI{10.1093/pnasnexus/pgae143}
\copyrightyear{2024}
\pubyear{2024}
\access{Advance Access Publication Date: Day Month Year}
\appnotes{Paper}

\firstpage{1}

\title[Astronomy’s climate emissions: global travel to scientific meetings in 2019]{Astronomy’s climate emissions: global travel to scientific meetings in 2019}

\author[a,b,c,$\ast$]{Andrea Gokus}
\author[d]{Knud Jahnke}
\author[e]{Paul M.\ Woods}
\author[f,g]{Vanessa A.\ Moss}
\author[h]{Volker Ossenkopf-Okada}
\author[i]{Elena Sacchi}
\author[j]{Adam R.\ H.\ Stevens}
\author[k]{Leonard Burtscher}
\author[l]{Cenk Kayhan}
\author[m,n]{Hannah Dalgleish}
\author[o]{Victoria Grinberg}
\author[p]{Travis A.\ Rector}
\author[q]{Jan Rybizki}
\author[r]{Jacob White}

\authormark{Gokus et al.}

\address[a]{Department of Physics \& McDonnell Center for the Space Sciences, Washington University in St. Louis, One Brookings Drive, St. Louis, MO 63130, USA}
\address[b]{Dr. Karl Remeis Sternwarte \& ECAP, Universit\"at Erlangen-N\"urnberg, Sternwartstr. 7, 96049 Bamberg, Germany}
\address[c]{Institut f\"ur Theoretische Physik und Astrophysik, Universit\"at W\"urzburg, Emil-Fischer-Straße 31, 97074 W\"urzburg, Germany}
\address[d]{Max-Planck-Institut für Astronomie, K\"onigstuhl 17, 69117 Heidelberg, Germany}
\address[e]{Independent researcher, London, UK}
\address[f]{ATNF, CSIRO Space and Astronomy, PO Box 76, Epping, NSW 1710, Australia}
\address[g]{Sydney Institute for Astronomy, School of Physics A28, University of Sydney, NSW 2006, Australia}
\address[h]{Universit\"at zu K\"oln, I. Physikalisches Institut, Z\"ulpicher Str. 77, 50937 K\"oln, Germany}
\address[i]{Leibniz-Institut für Astrophysik Potsdam (AIP), An der Sternwarte 16, 14482 Potsdam, Germany}
\address[j]{International Centre for Radio Astronomy Research, The University of Western Australia, Crawley, WA 6009, Australia}
\address[k]{Leiden Observatory, PO Box 9513, 2300 RA, Leiden, The Netherlands}
\address[l]{Department of Astronomy and Space Sciences, Faculty of Science, Erciyes University, 38030 Melikgazi, Kayseri, T\"urkiye}
\address[m]{Politics \& International Relations, University of Southampton, University Rd, Southampton, SO17 1BJ, UK}
\address[n]{Department of Physics, University of Oxford, Keble Rd, Oxford, OX1 3RH, UK}
\address[o]{European Space Agency (ESA), European Space Research and Technology Centre (ESTEC), Keplerlaan 1, 2201 AZ Noordwijk, The Netherlands}
\address[p]{University of Alaska Anchorage, 3211 Providence Dr., Anchorage, AK, 99508 \ USA}
\address[q]{Weimar Institute of Applied Construction Research, \"Uber der Nonnenwiese 1, 99428 Weimar, Germany}
\address[r]{Independent researcher, Vancouver, Canada}

\corresp[$\ast$]{To whom correspondence should be addressed:\href{email:dr.andrea.gokus@gmail.com}{dr.andrea.gokus@gmail.com}}

\received{Date}{0}{Year}
\accepted{Date}{0}{Year}


\abstract{Travel to academic conferences --- where international flights are the norm --- 
is responsible for a sizeable fraction of the greenhouse gas (GHG) emissions 
associated with academic work. In order to provide a benchmark for comparison 
with other fields, as well as for future reduction strategies and assessments, we 
estimate the CO$_2$-equivalent emissions for conference travel in the field of 
astronomy for the pre-pandemic year 2019. The GHG emission of the international astronomical community's \totMeetings~conferences and schools in 2019 amounted to 
\totEmissionsRounded, assuming a radiative-forcing index (RFI) factor of 1.95 for air travel. 
This equates to an average of \emissionsPerCapMeetings\ per participant per meeting.
The total travel distance adds up to roughly 1.5 Astronomical Units, that is, 1.5 times the distance between the Earth and the Sun.
We present scenarios for the reduction of this value, for instance with virtual conferencing or hub models, while still prioritising the benefits conferences bring to the scientific community.}

\boxedtext{
The climate crisis is the biggest challenge of our lifetime, and systemic changes are needed to reduce greenhouse gas (GHG) emissions. Professional travel, especially international flights, accounts for a significant portion of the carbon footprint of academia. In our study, we take a look at air travel from scientific astronomy meetings, held in the pre-pandemic year 2019, to obtain quantitative information from the international astronomical community that can be used to set realistic targets for the necessary emission reduction. We explore alternative scenarios for meetings, which lower the amount of air travel, and thereby the carbon footprint of the scientific community, but still include or even increase the benefit of meetings for international collaboration.
}

\maketitle

\section{Introduction}

There is unequivocal scientific evidence that the current climate change on Earth is caused by the emission of anthropogenic greenhouse gases (GHGs), dominated by CO$_2$ and CH$_4$ \citep{ipcc21ch3}.
Global warming leads to an increase in the frequency and severity of extreme weather events including droughts, flooding, and (large) wildfires — and adds extreme costs and health risks to humans \citep{ipcc_chapt11}.
In order to limit the effects of the accelerating climate crisis for life on Earth, it is imperative to drastically reduce the emission of GHGs.
Most GHG emissions are associated with fossil fuel use, $\sim$16\% of the global GHG are from the transport sector, due to mostly running on fossil fuels. With all effects included, aviation contributes 7\% to the global human-made climate forcing across all sectors, with 2--3\% of this directly from CO$_2$ warming \citep{sheet2019climate, international2020tracking}, the rest from other effects, which include mechanisms driven by airplane-created oxides of nitrogen, soot, oxidised sulphur species, and water vapour. The latter contribution creates an impact through contrails that act similar to high cirrus clouds, which have a net warming effect by trapping the Earth's thermal infrared radiation \citep{avila2019}. Aviation is hence one of the major drivers of global climate change, both in the transport sector but also globally\footnote{`Updated analysis of the non-CO$_2$ effects of aviation' by the EU Aviation Safety Agency: \url{https://climate.ec.europa.eu/news-your-voice/news/updated-analysis-non-co2-effects-aviation-2020-11-24\_en}}\citep{lee2021}. 

Travel in academia is common practice. Researchers benefit from networking with their colleagues during meetings \citep{Oester2017}, and a correlation of the visibility of a scientist with their amount of air travel seems to exist \citep{Berne2022}.
However, in light of the climate crisis, researchers, particularly working in the field of medicine, ecology, biology, and astronomy, have started to critically assess their own travel behaviour by calculating GHG emissions related to individual, large, annual meetings \citep{Callister2007, Yakar2020, Bousema2020, Passalacque2021, Wortzel2021, Periyasamy2022, Mcclintic2023}. 
In general, airtravel to conferences makes up the largest contribution of GHG emissions to a meeting by far \citep{Periyasamy2022,Mcclintic2023}.
Some studies also compare how the carbon footprint of a recurring meeting evolved over time \citep{Ponette2011, Jaeckle2019, Milford2021}, or how an in-person conference compares to a virtual one \citep{West2022}.
For a more comprehensive overlook, we refer the interested reader to the recently released book titled `Academic Flying and the Means of Communication' \citep{academicflying_book}, which presents studies connected to academic travel extending beyond conference attendance and discusses how to decarbonise the travel culture within academia.

The academic research field of astronomy and astrophysics is an international discipline.
The International Astronomical Union (IAU) counts close to 12,000 active members (with a historical peak of $\sim$14,000 members in 2020) from $\sim$90 nations%
\footnote{IAU Member statistics: \url{https://www.iau.org/public/themes/member_statistics/}}. 
The American Astronomical Society (AAS) has 7,000 members, of which more than 700 reside in almost 60 countries outside the USA \citep{aas2020}.
A study \citep{knodlseder2022} estimates a total number of $\sim$30,000 professional researchers in astronomy and astrophysics, including graduate students.
Historically, due to this widespread distribution, the need for data collection from remote observatories, as well as scientific exchange, astronomy developed a strong dependence on international travel.
Reasons for travel include observing at telescopes, project meetings, research visits, the installation of new instruments and telescopes, and the dissemination of knowledge and learning at conferences, workshops, schools and outreach events. The large majority of the distances travelled by astronomers is related to the attendance of conferences, workshops, and other meetings \citep{Blanchard2022}.

While geographically diverse, astronomy research is a comparatively small community. 
Globally, $\sim$8 million researchers work in different fields of science{\footnote{UNESCO science report: \url{https://en.unesco.org/unesco_science_report/figures}}}, meaning astronomers make up a fraction of $<$1\% of the world's scientists. Nonetheless, numbers for France \citep{Blanchard2022} position astronomy as the science discipline with the largest travel climate impact per researcher.
This travel comes at a cost aside from the obvious financial one:~a substantial immediate emission of GHGs. 

However, these travel emissions are not the sole source of astronomy's carbon footprint. In recent years, a number of studies calculated the emissions for different aspects of astronomy, providing quantitative information for GHG emissions of specific elements of the community.
These studies provided reasonable estimates for GHG emissions within the Australian astronomy community \citep{stevens2020}, the Dutch astronomy community \citep{vdTak2021}, research institutes \citep{jahnke2020,martin2022}, space missions \citep{knodlseder2022} as well as ground-based observatories \citep{flagey2020a,flagey2020b,eso2020,grand2021,knodlseder2022,mccann2022}, and conducted representative calculations of co-located and online conferences \citep{burtscher2020}. 
These data revealed work-related mean emissions per researcher and year that, depending on the country and institute, range between 4.7\,\tceq~\citep{vdTak2021} and 41.8\,\tceq~\citep{stevens2020}, all of which are beyond what is sustainable for the field of astronomy, and incompatible with any emission limits necessary to fulfil the Paris Agreement\footnote{United Nations Climate Change (UNFCCC) secretariat -- The Paris Agreement: \url{https://unfccc.int/process-and-meetings/the-paris-agreement}} \citep{sru2020}. 

As a next step for the assessment of astronomy-related GHG emissions and to form a basis for reduction measures, we quantitatively evaluate a significant source of emissions: those related to travel to astronomy conferences and schools.
We take stock of globally all major astronomy meetings for the pre-pandemic year 2019 as an ‘old normal’ baseline, and investigate where these meetings take place, where participants come from, and what the financial, time, and climate costs are. We also consider how this impacts both equity and sustainability responsibilities in the field.
While we discuss potential measures, and approaches to reduce or eliminate flight-related GHGs of astronomy, we do not purport to immediately ‘solve’ the issues identified, as this is a task requiring input from the whole community.  
We have conducted this study primarily to contribute a quantitative data set on astronomy travel, which can be used to subsequently derive reduction needs, options, and potential focus points of action. While concretely assessing the field of astronomy, the findings can be mapped onto any other scientific fields with similar structure regarding conferences and schools --- this study is in principle investigating fundamental structures of science as a whole. As part of our discussion we also briefly address how changing some of our conference habits might have potential for improved equity and inclusivity in the field of astronomy.

\section{Results}

We compiled a data set of \totMeetings~meetings (conferences and schools) that took place around the globe between January 1 and December 31, 2019, and were open to attend by anyone in the astronomical community. 
We compute an amount of \totEmissionsRounded~for the total travel to those \totMeetings~scientific astronomy meetings in 2019.
Substantial research makes us confident that the derived list of meetings is (as near as can be) complete. Hence this study represents the whole of global astronomy. Closed-door collaboration meetings were not considered for this study, as they are hard to assess from the outside --- they require a future study of their own. The process of compiling the meeting list is described in the `Materials and Methods' section of this paper.
Among the open meetings, we were able to retrieve information about the number of participants and their associated home institutions for exactly \totGoodMeetings. 
The remaining \totColdCaseMeetings, for which we could obtain only incomplete data, are incorporated into overall numbers with statistical arguments later in the paper.

We found that astronomy meetings fall into two broad categories, which we refer to as `conferences' and `schools'. Conferences are predominantly aimed at the entire astronomical community, with the primary purpose of conveying scientific findings and a delivery of presentations as talks or posters. Schools are usually aimed at students or early-career scientists, and content is mainly presented in the form of lectures, some including hands-on activities. Of the \totGoodMeetings~meetings that we obtained a complete data set for, we counted \totGoodConf\ conferences and \totGoodSchool\ schools. Based on these data, we estimate the travel-related carbon footprint for each meeting individually, and then combine all meetings to assess the climate impact of astronomy's pre-pandemic meeting habits. 

\begin{figure*}
    \centering
    \includegraphics[width=.9\textwidth]{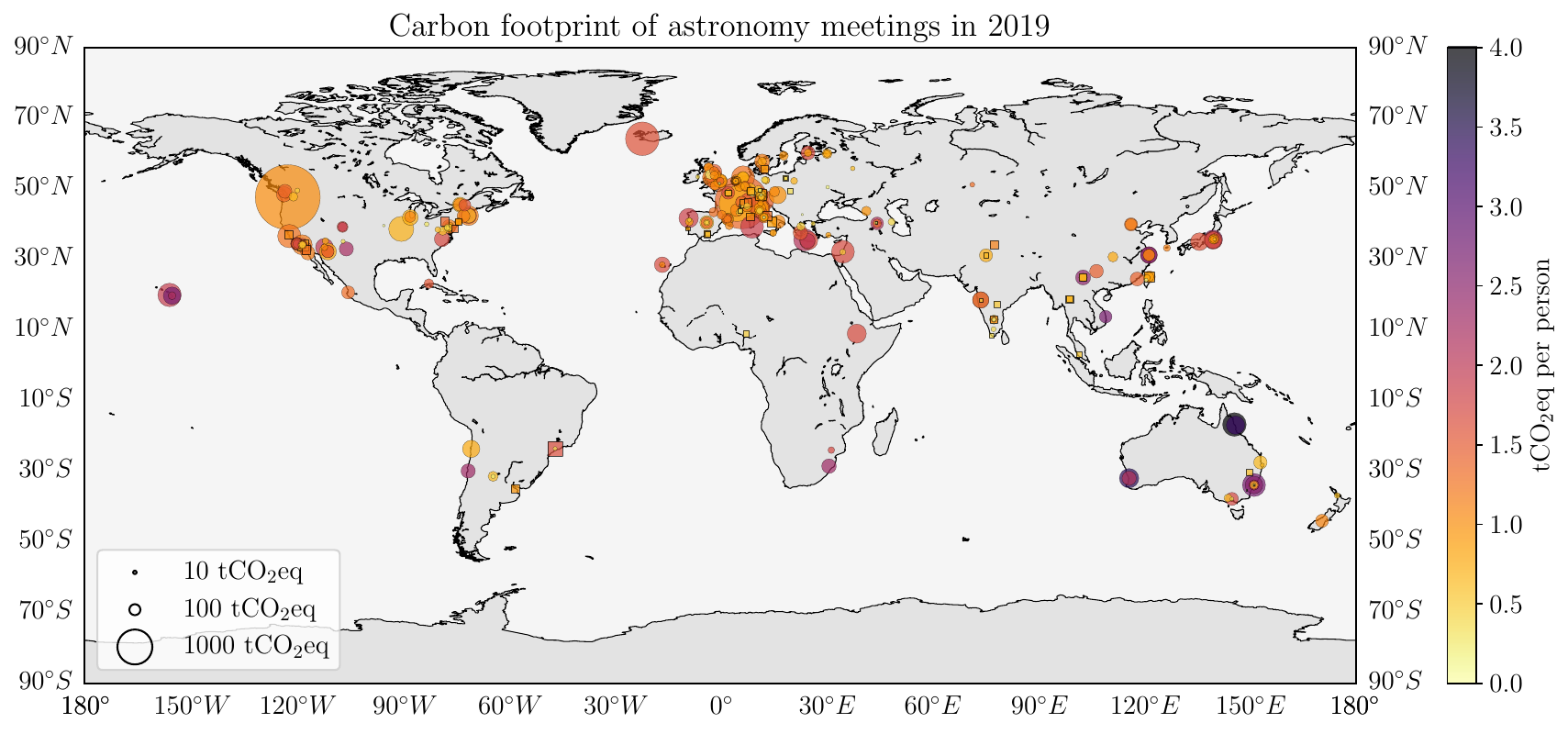}
    \caption{Distribution of all 2019 astronomy/astrophysics meetings around the globe. Conferences are shown as circles and schools are shown as squares. The size of each marker corresponds to the overall amount of GHG emissions related to travel to each meeting, while the color-scale indicates the mean emission per participant for each meeting. A darker color implies a higher carbon footprint per person, which is related to travel over larger distances.}
    \label{fig:map_of_meetings}
\end{figure*}

\begin{figure*}
\centering
\includegraphics[width=0.95\textwidth]{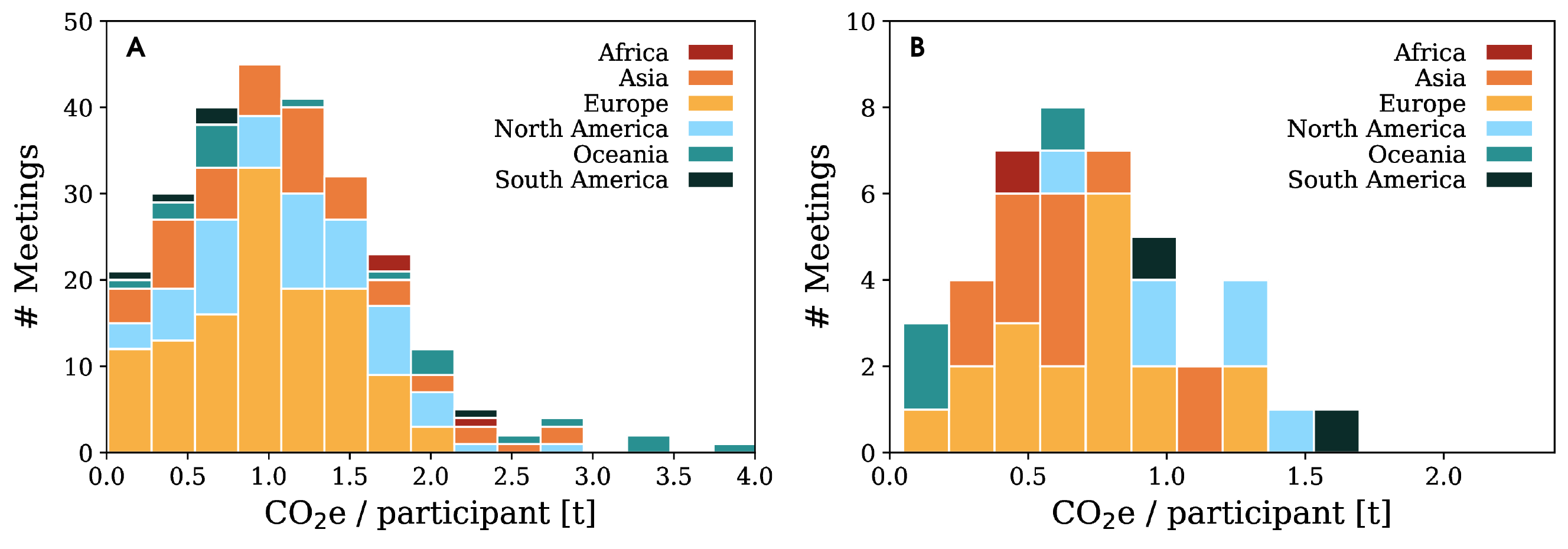}
    \caption{Stacked distribution of the mean travel emissions per participant for conferences (panel A) and schools (panel B) on different continents. Colours indicate the continent on which a meeting took place. The stacking order is set by the number of meetings per continent from large to small, i.e., Europe - North America - Asia - Oceania - South America - Africa for conferences, and Europe - Asia - North America - Oceania - South America - Africa for schools.}
    \label{fig:histogram_emissionpercapita}
\end{figure*}

The locations of all \totGoodMeetings~meetings are shown on the map in Fig.~\ref{fig:map_of_meetings}, with conferences indicated by circles and schools by squares. In addition to the geographical information, we include the meeting's total carbon footprint by the size of the marker, and the average carbon footprint per participant via the color-scale\footnote{Emission numbers assume RFI of 1.9--2.0 for air travel, see `Materials \& Methods' for details.}.
For these \totGoodMeetings\ meetings we find that the total amount of GHG emissions is \totGoodEmissions, and the added-up travelled distance is equivalent to $\sim$5650 times the circumference of the Earth, or $\sim$\,1.5 Astronomical Units, i.e.,\ the distance from Earth to the Sun.
Other studies which assessed single, large conferences, obtain similar or even higher values for the average travel-related GHG emissions per meeting participants \citep{Callister2007, Bousema2020, kloewer2020, Yakar2020, Wortzel2021}.
The resulting distribution of the mean travel emissions per participant for conferences and schools are shown in Fig.~\ref{fig:histogram_emissionpercapita}, with an additional distinction based on the continent of a conference venue. The statistical properties are listed in Table~\ref{tab:general_statistics}.

%

%

%
\begin{table*}
\caption{Statistics of evaluated emissions for \totGoodConf~astronomy/astrophysics conferences and \totGoodSchool\ astronomy/astrophysics schools in 2019, listing global numbers and statistics for each continent. From left to right: total number of meetings -- accumulated travel route for all meetings -- total amount of travel-related CO$_2$e emission -- average number of participants per meeting -- average travel-related carbon emissions per meeting -- the mean fraction of local attendees. Uncertainties correspond to the standard deviation, except for the number of conference participants at European, North- and Central American, and all worldwide conferences, for which we give the upper and lower 68\% uncertainty as those follow a positively skewed log-normal distribution.
We consider a participant to be local if the round-trip to the meeting location is $\leq$200\,km.
Note that we could only analyse a very little amount of schools with venues in Africa, Oceania, and South America, and therefore no statistical conclusions can be drawn for those continents.}
\centering
\begin{tabular}{l|cccccccc}
\hline\hline
Conference venue & Number of & Cumulative travelled  & Cumulative CO$_2$e  & Mean number & Mean emission & Mean local \\
location  & meetings  & distance [km] & emissions [t] & of participants & per participant {[}\tceq{]} & participants [\%]   \\
\hline
\multicolumn{7}{c}{CONFERENCES}\\
\hline
World-wide  & \totGoodConf & $2.16\times10^8$  & 36254  & $128^{+33}_{-86}$   & $1.1\pm0.6$  &  $21\pm18$  \\
\hline
Africa   & 3   & $3.13\times10^6$  & 515  & $93\pm80$  & $1.9\pm0.4$  & $9\pm8$\\
Asia   & 49   & $3.10\times10^7$ & 5125   & $91\pm54$   & $1.1\pm0.7$  & $26\pm20$ \\
Europe   & 124 & $9.76\times10^7$  & 16315  & $127^{+37}_{-87}$   & $1.0\pm0.5$  & $17\pm14$\\
North America & 59  & $6.50\times10^7$  & 11074  & $176^{+64}_{-129}$   & $1.2\pm0.6$ & $22\pm18$\\
Oceania  & 18  & $1.62\times10^7$  & 2694  & $83\pm55$  & $1.6\pm1.2$  & $24\pm21$\\
South America  & 5  & $3.08\times10^6$ & 531   & $132\pm115$   & $0.8\pm0.8$  & $40\pm28$  \\
\hline
\multicolumn{7}{c}{SCHOOLS}\\
\hline
World-wide  & \totGoodSchool & $1.05\times10^7$  & 1786  & $54\pm27$  & $0.7\pm0.4$  &  $23\pm21$ \\
\hline
Africa   & 1   & $1.99\times10^5$  & 33  & $70$  & $0.46$  & $16$ \\
Asia   & 12   & $2.48\times10^6$ & 435   & $53\pm21$ & $0.65\pm0.27$ & $25\pm22$ \\
Europe   & 18 & $4.01\times10^6$  & 690  & $53\pm24$   & $0.69\pm0.33$  & $20\pm17$ \\
North America & 6  & $2.02\times10^6$ & 340  & $59\pm45$   & $1.10\pm0.28$  & $20\pm17$  \\
Oceania  & 3  & $2.10\times10^5$  & 36  & $32\pm21$  & $0.23^{+0.35}_{-0.18}$ & $49\pm41$  \\
South America  & 2  & $1.53\times10^6$  & 252   & $87\pm42$   & $1.3\pm0.5$  & $32\pm14$  \\
\hline
\end{tabular}
\label{tab:general_statistics}
\end{table*}


\subsection{Mean travel emissions}
The mean, global round-trip travel emission to a single conference (see Table~\ref{tab:general_statistics}) is \emissionsPerCapConf\ per participant -- for context, in 2019 the average per-capita carbon footprint in a developed country was 9.5\,\tceq~while in, e.g., Africa, it was 1.2\,\tceq~\citep{ipcc_tecsum}.
In order to keep the climate warming below 2$^\circ$\,C, the world-wide average needs to be below 3.3\,\tceq\ by 2030 \citep{essd-10-2141-2018}.
The mean emissions per participant and meeting vary slightly depending on the geographic location of meetings.
For conferences, the highest and lowest mean emissions per participant are attributed to meetings taking place on the continents of Africa and South America, respectively, but the number of meetings that took place in these regions was small. 
Meetings in Europe are likely to have a slightly lower carbon footprint in terms of mean per-person travel emissions, as journeys of a Eurocentric audience are shorter compared to North America and Asia, and longer distances can be travelled by train compared to, e.g., the USA. Interestingly, European meetings tend to have the second-lowest participation by local participants, even though these meetings have the second-lowest mean emissions.

Compared to conferences, schools have a lower carbon footprint, which is \emissionsPerCapSchool~per participant.
Schools in Asia appear to have slightly more local participants and produce fewer GHG emissions on average than similar meetings in Europe and North America. The carbon footprint of European schools is lower than that of European conferences, while for meetings in North America, schools and conferences are very similar in terms of average travel emissions per participant.
For the other regions, we could only obtain data for one school in Africa, two in Oceania, and three in South America, which does not allow us to draw any conclusions for these locations due to insufficient statistics. 

Three meetings exceeded average travel emissions of 3.0~\tCOTe\ per person, and all of them took place in Australia. Because Oceania is geographically disadvantaged by being far from many established centers of astronomical research, and thus from where the bulk of astronomers travel from, international meetings held there tend to have higher mean travel emissions per participant. In addition, two of these meetings had no local participants, as the conferences took place in remote locations.

\subsection{Localness}
\begin{figure}
    \centering
    \includegraphics[width=.45\textwidth] {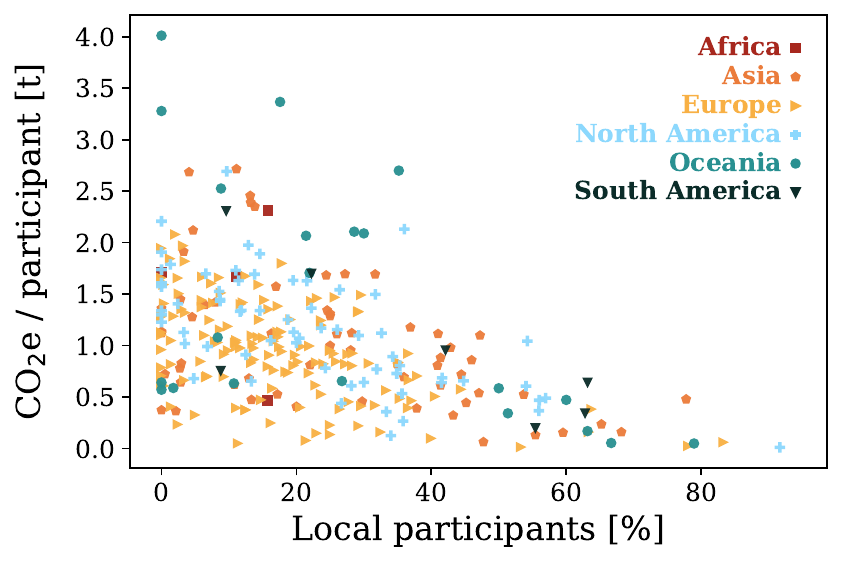}
    \caption{Localness vs.\ mean travel emissions per participants for the conferences. Colours and markers indicate the continent on which a meeting took place.}
    \label{fig:localness_vs_emissions}
\end{figure}

We considered a participant to be a local if the round-trip to the meeting location is $\leq$200\,km.
It is shown in Fig.~\ref{fig:localness_vs_emissions} that meetings with a higher fraction of locals have significantly lower mean emissions per participant than meetings with few locals. However, the distribution of the carbon footprint per participant as a function of localness scatters significantly in the range between 0\% and 35\% of local participants, indicating that a small number of local participants does not necessarily imply high average travel emissions. 
This scattering is observed for meetings on all continents.
However, as air-travel emissions scale with distance, small mean emissions per participant correspond to institutions of origin that are on average not too far away. As we discuss later, minimising the total distance traveled is a good approach to reducing conference emissions.

%

\subsection{Meeting sizes}
\begin{figure}
    \centering
    \includegraphics[width=.45\textwidth]{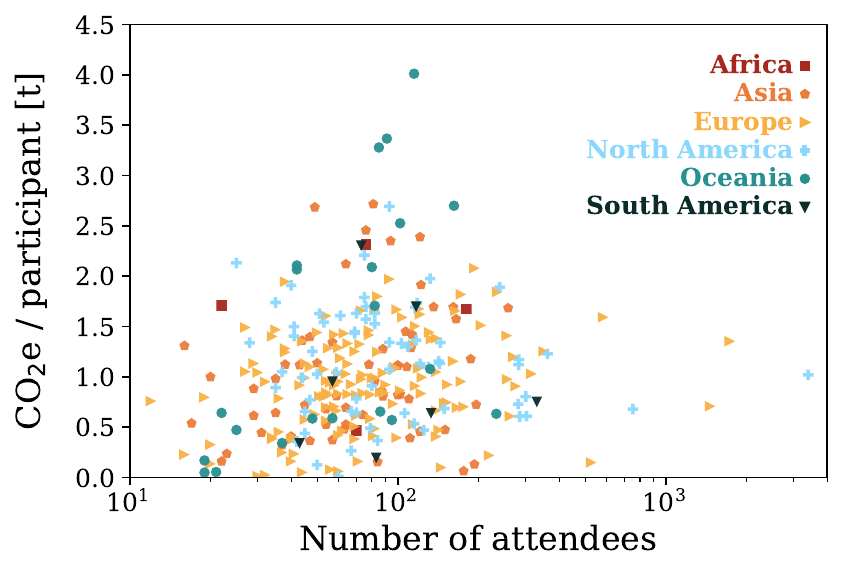}
    \caption{Illustration of participant numbers versus the \COT~emission per participant. Colours indicate the continent on which a meeting took place.}
    \label{fig:correlation_numbers_emissionpercapita}
\end{figure}

The number of participants for the \totGoodConf~analysed conferences varies over two orders of magnitude, with a majority of the number of participants
being in the range of 128$^{+33}_{-86}$ for conferences. Four of them counted more than 700 participants, namely the 233rd and 234th American Astronomical Society (AAS) meetings, the joint meeting between the Europlanet Science Congress (EPSC) and the Division of Planetary Science (DPS) within the AAS, and the annual European Astronomical Society (EAS) Meeting. The 233rd AAS meeting was the largest conference with 3,396 participants. 
The mean number of participants at schools is significantly lower with an average of 54$\pm$27 participants.
Figure~\ref{fig:correlation_numbers_emissionpercapita} shows, for all meetings, the number of participants versus the average \COT\ emissions per participant, and shows that the mean travel emission per participant is independent of the size of a meeting.
Among the four meetings with more than 700 participants, all of them were aimed particularly at the American and European Astronomy communities (233rd and 234th AAS meeting, EAS meeting, and a joint meeting of both the American and European planetary science division of the AAS and EPSC).
%

\subsection{Meeting locations}
\begin{figure}
    \centering
    \includegraphics[width=.49\textwidth]{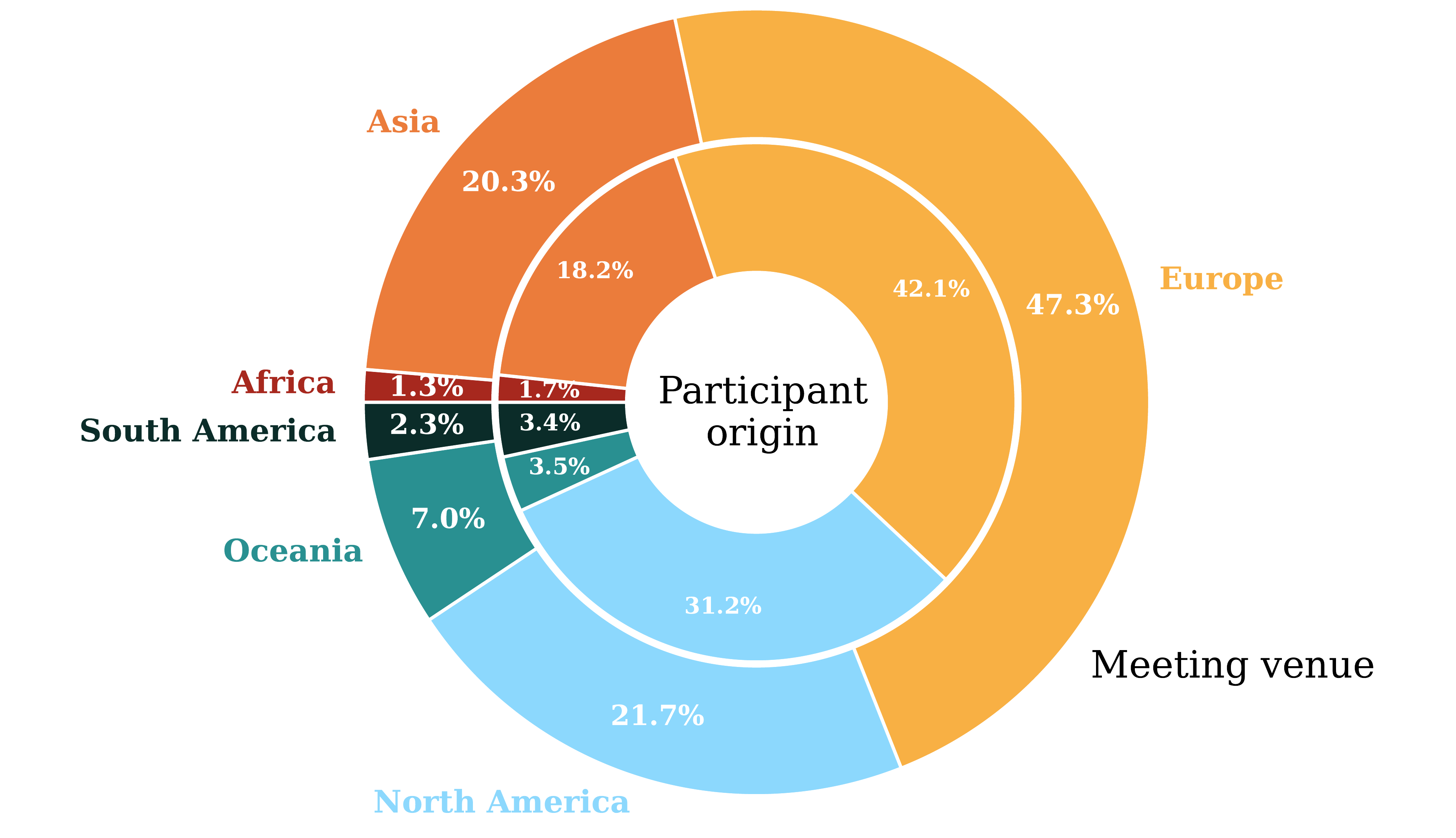}
    \caption{Geographical distribution of the meeting venues by continent for the \totGoodMeetings\ meetings with complete participant data (outer ring), and geographical distribution of the home institutions of all participants (inner ring). With regards to meeting venues and participants, Europe and North America dominate both distributions. Note: Individual scientists will be counted multiple times if participating in multiple meetings.}
    \label{fig:donut_venue_participants}
\end{figure}

We visualise where meetings take place and where participants travel from in Fig.~\ref{fig:donut_venue_participants} (see also Supplementary Table 1 for total numbers of conferences and schools hosted in different countries). We find that nearly half (47.2\%) of all meetings take place in Europe, and roughly a fifth of the meetings take place in North America\footnote{Including the Caribbean, where one meeting was held in 2019.} (21.7\%) and Asia (20.4\%), respectively. The remaining meetings are distributed between Oceania (7\%), South America (2.3\%), and Africa (1.3\%).
The distribution in terms of meeting participation for the home institutes looks somewhat similar with European-based astronomers being the largest group (42.1\%), followed by their North American-based colleagues (31.2\%). We find that about a fifth of all meeting attendees originate from institutes in Asia (18.3\%), while the remaining $\sim$10\% is distributed among Oceania (3.5\%), Central \& South America\footnote{Note that no meetings in our analysis were hosted in Central America in 2019} (3.4\%), and Africa (1.7\%). 
If we compare the numbers of participants with the IAU member statistics\footnote{\url{https://www.iau.org/administration/membership/individual/distribution/}; IAU members per continent: Africa: 2.2\%; Asia: 22.3\%; Europe: 42.9\%; North America: 24.9\%; Oceania: 3\%; Central \& South America: 4.7\%}, we find that our distribution for meeting participants approximately matches the IAU member distribution for Europe, Africa, and Oceania. The fraction of IAU members at Asian institutions is higher than that seen participating in meetings, as is the case for astronomers located at Central \& South American institutes. It is possible that our compiled meeting list lacks some local events organised for these communities in particular.
The fraction of attendees from North America is significantly higher than in the IAU membership statistics. 
Since we track (anonymised) conference attendances and not individuals, this difference may be an indication that astronomers from North American institutes participate in more meetings than the average. It also might be a reflection of low IAU membership in North America (e.g., because membership of just the American Astronomical Society is preferred).
We also note that the IAU membership requires a PhD and therefore, those statistics exclude undergraduate and graduate students, who are also taking part at conferences, and make up the majority of participants at schools. Due to the intentional anonymisation during our data collection process a further data-driven analysis, e.g., regarding career status, gender, or other characteristics is not possible.

%

\begin{figure*}
    \centering
    \includegraphics[width=.95\textwidth]{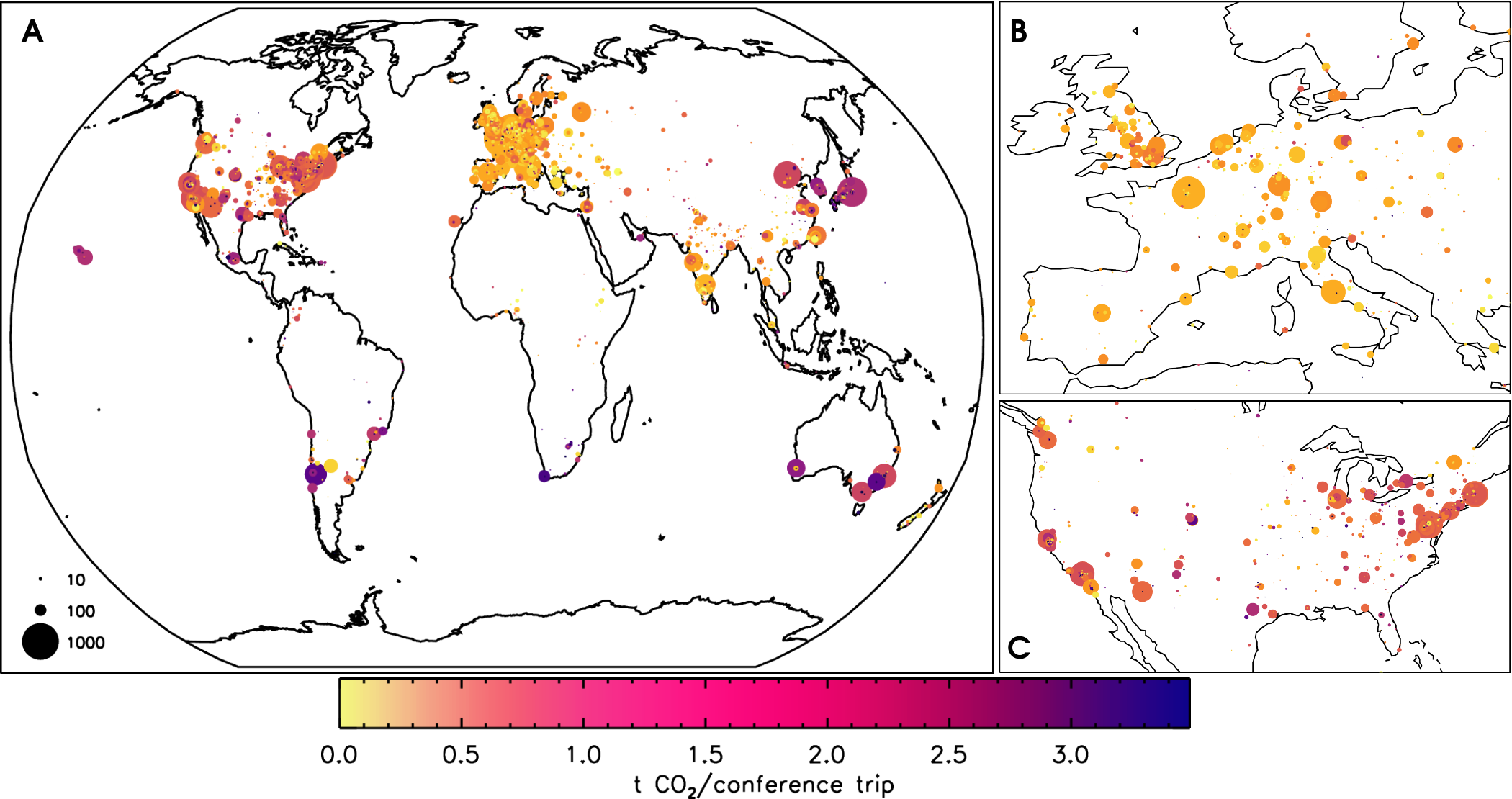}
    \caption{Global distribution of travel origins of conference and school participants (panel A).
    The area of the circles at each location represents the number of trips taken from there. The colour gives the average \COT\ intensity of individual trips per location. Zooms to Europe and North America with rescaled symbols are shown in panels B+C for better visibility.}
    \label{fig:map_visitor-places}
\end{figure*}

Figure~\ref{fig:map_visitor-places} illustrates the location of the home institutions of conference and school participants in more detail. We scale the size of the circles by the number of trips taken from each location. It is expected that this correlates to the amount of scientists working at an institute in this region. The colour-coding indicates the emissions per conference travel averaged over all trips from that location. It can be seen that the most trips are taken from locations in Europe and the east coast of the USA. The amount of trips originating from China, Japan, and Australia are comparable to those from the west coast of the USA. As with the meeting venues, we see the effect that researchers in Europe profit from the on-average smaller distances allowing for more trips at a lower \COT\ emission while researchers from South Africa, Hawai'i, Japan or Australia have little chance to reduce their carbon footprint when participating in in-person international meetings. 

%

\subsection{Assessment of emissions from meetings without participant information}
To estimate the missing emissions from the \totColdCaseMeetings\ meetings for which we could not retrieve any data on the home institutions of participants, we use the average emissions per participant and continent, which we derived from the other meetings. Depending on whether we find information on the number of participants, e.g., number of people in the group picture of the meeting, or the number of names listed in a program, we use either those, or estimate the number of participants by the topic and venue of the meeting.
For the remaining meetings that do not show any indication for number of participants, we take the average number of participants per continent as listed in Table~\ref{tab:general_statistics}. 
From this, we derive an additional amount of \totColdCaseEmisisons, with \totColdCaseConfEmissions~resulting from conferences, and \totColdCaseSchoolEmissions~from schools. For an estimate of the uncertainty, we take into account the derived standard deviations of the average emission per participant and continent. The continent-specific numbers for conferences and schools are listed in Table~\ref{tab:coldcases}, and the details for all meetings without data on participants' home institutes are given in the Supplementary Material in Tables S3 and S4 for the conferences and schools, respectively.
In total, we calculate travel emissions of \totEmissions~due to identified meetings in astronomy and astrophysics in 2019. Here the uncertainty of $\sim$1\% reflects the overall uncertainty in the number of participants and their travel origins. In the section on systematic uncertainties (Sect.~\ref{sec:systematic_unc}) we discuss how, e.g., the choice of RFI might be adapted for different further uses of this number. With this in mind, we round the calculated travel emissions to \totEmissionsRounded.

\begin{table}[t]
\caption{Statistics for estimated emissions for \totColdCaseConf~astronomy/astrophysics conferences and \totColdCaseSchool~astronomy/astrophysics schools in 2019, for which full participant data was not available. If the number of participants could not be estimated via, e.g., the conference photo, we assigned a meeting the continental mean value presented in Table~\ref{tab:general_statistics}. To compute the total \COT~emissions, we use the continent-specific mean emission per capita. The uncertainties on the total \COT~emissions have been derived via standard error propagation, and were derived individually for the continent-specific locations and the entire world-wide sample.}
\centering
\begin{tabular}{l|cccccc}
\hline\hline
Conference venue & Number of & CO$_2$e  & Participants \\
location  & meetings  & emission [t] & Mean \\
\hline
\multicolumn{4}{c}{CONFERENCES}\\
\hline
World-wide  & \totColdCaseConf & $4154\pm349$ & $79\pm44$ \\
\hline
Asia & 11 & $1062\pm205$ & $88\pm11$\\
Europe & 19 &  $1611\pm164$ & $85\pm41$ \\
North America & 15 & $1267\pm182$ & $70\pm63$ \\
Oceania & 2 & $158\pm99$ & $50\pm43$ \\
South America & 2 & $156\pm74$ & $65\pm7$\\
\hline
\multicolumn{4}{c}{SCHOOLS}\\
\hline
World-wide  & \totColdCaseSchool & $400\pm53$  & $42\pm14$ \\
\hline
Asia & 5 & $100\pm20$ & $31\pm8$ \\
Europe & 8 & $266\pm46$ & $48\pm13$\\
\hline
\end{tabular}
\label{tab:coldcases}
\end{table}


\section{Discussion}
At large, researchers across all fields are concerned about the worsening climate crisis, and the willingness to reduce GHG emissions exists among academic staff \citep{thaller2021}. Here, we want to address different options that will allow to reduce the travel-related carbon footprint of scientific meetings in order to give incentives for future meeting planning. 
We want to point out that carbon offsets, albeit seeming like an easy solution to claim a net reduction of GHG emissions, are no alternative to reduce travel emissions \citep[e.g.,][]{calel2021, anderson2012, bullock2009}.
This discussion is led by the assumption that the frequency and size of meetings represents the fundamental communication needs of the field -- so, we do not propose to simply stop having meetings, as such. While, to some degree, a fraction of conferences could probably be cancelled without overall loss of communication, this is likely not a feasible solution as a whole. Instead, we discuss options for both reducing climate emissions for in-person meetings, as well as possibilities for hybrid or fully online meetings \citep[see also][]{allea2022}. Such a shift in our way of communication will have to be carried by the community: from the will to optimize meeting locations, to consider new advances in technology for hybrid meetings and immersive online options, and to re-evaluate the ``career prestige'' of organizing or participating in any given meeting format.

\subsection{Meeting locations}
Traditionally, conference organisers set a location for their conference based on (i) where they themselves are situated, (ii) a particular destination's benefits (retreat-based conferences), (iii) a set rotation (e.g., AAS meetings; national astronomy meetings), and/or (iv) exotic locations, typically vacation spots, that are chosen to encourage spending some time there before or after the meeting. The origin of travel for most conference participants is rarely known precisely in advance of setting the conference location, particularly for general conferences that cover a variety of astronomical topics. However, for collaboration-based meetings or subject-specific meetings, one often knows in advance where the majority of participants will travel from. This \emph{a priori} knowledge could be utilised to reduce conference-based emissions significantly, or alternatively expressions of interest in a conference could be sourced from the community prior to selecting a physical location. 

Taking for example the meeting which resulted in the highest emissions per participant in 2019, a conference based in the tropical Australian village of Palm Cove, Queensland, attracted 115 participants, of which 105 travelled from outside Australia, generating $\sim$461\,\tceq. Holding this meeting in mainland Europe (e.g., Heidelberg, Germany) or northeastern USA (e.g., Washington, D.C.) could have reduced the \COT~emissions by more than half. 
Of course, there are a number of reasons that might motivate hosting a conference in such a location: first of all, many regions in the world will always be geographically disadvantaged regarding the main global concentrations of astronomy research, i.e.\ Europe and North America (e.g., Fig.~\ref{fig:map_visitor-places}), and it benefits the international community to increase access to knowledge dissemination and interaction beyond established hubs of astronomers. 
Secondly, scenic attractions and cultural immersion can be utilised as an additional incentive for researchers to embark on long trips to distant locations, while bringing economic benefits to local communities. Such reasons noted, we urge that in
these times of strong necessity to contain the climate crisis, more attention should be paid to the environmental impact of travel when choosing a conference location, weighing up the costs against the benefits and ensuring that decisions about in-person formats can be justified. One should also take into consideration the modes and duration of travel, and look for opportunities to lengthen or increase the value of a trip that may otherwise be quite high in carbon cost.

In general, long-distance travel contributes considerably to the climate crisis \citep{VANGOEVERDEN201628}.
Following the classification of the International Air Transport Association (IATA)\footnote{Air Transport \& Travel Industry -- IATA EDIFACT AND XML CODESET: \url{https://standards.atlassian.net/wiki/download/attachments/607420437/IATA\%20Code\%20set\%20Directory.pdf}}, 
we can identify four categories of journeys, depending on the length of flight time: short-haul (SH; 3 hrs or less), medium-haul (MH; 3--6\,hrs), long-haul (LH; 6--16\,hrs) and ultra-long-haul (ULH, more than 16\,hrs). Given that a 3-hour flight covers more distance than the longest train journey we assumed (1,000\,km), we consider all train journeys to be SH journeys, too, even though the actual travel time is more than 3\,hrs. We note that while the classification is solely based on the duration of the flights, air travel requires an additional time commitment for travel to the airport, security procedures, and arrival at the gate 30\,minutes or more prior to the departure of a flight. Depending on the rail network, train travel can therefore be of similar duration as the total amount of time necessary to travel short-haul. Details on the assumptions for different continents are given in the section on calculating \COT~emissions in `Materials and Methods' (Sect.~\ref{sec:matmet}).

Comparing all the trips to the \totGoodMeetings\ meetings, split into these three categories, with the \COT\ emissions, reveals that LH flights make up more than 60\% of emissions while only accounting for 21\% of all trips (see Figure S2 in Supplementary Material). This comes as no surprise, but strengthens the argument for different meeting concepts (c.f., \citep{moss2021_na,neubauer2021,lowell2022}) in order to mitigate travel-related emissions.
We emphasise that minimising the GHG emissions should become a more prevalent factor in determining where any meeting should be held. Furthermore, we note that the appropriate meeting format should be chosen upon critical consideration, based on the needs and expected outcomes of a given meeting.

\subsection{Alternative meeting concepts}
Instead of the traditional meeting concept of gathering physically in just one place, we consider three alternatives to estimate the potential for saving travel-related \COT\ emissions: hubs, where participants travel to the nearest of several locations that are connected virtually; hybrid meetings, which involve an in-person component and a virtual-attendance component, ideally on equitable footing; and online-only meetings where all activities, social events included, are hosted virtually. Such alternative meeting formats can make long-distance flights unnecessary, which significantly contribute to a meeting's climate impact.

We want to quantitatively consider the transformation of a single-venue meeting to a hub model. 
For example, the 233rd AAS meeting in Seattle, WA (USA) --- the largest astronomy-related conference in 2019 with 3,396 participants --- generated 3,462\,\tceq. If located more centrally, however, (e.g. Minneapolis, MN) the meeting would have generated 2,615\,\tceq\ --- 25\% less. Alternatively, we can consider 2, 3, and 4 hub locations, with each participant travelling to the nearest hub:
\begin{itemize}
\item 2-hub model. We divide the 48 US mainland states into east and west, with the north--south dividing line close to Dallas, TX. Western participants meet in Los Angeles, CA, eastern participants in Baltimore, MD. The western contingent expends 692\,\tceq, the eastern 685\,\tceq. Combined emissions decrease by 60\% compared to everyone travelling to Seattle.   
\item 3-hub model. This scenario adds a European hub to the existing Los Angeles and Baltimore hubs, reducing the impact of the Baltimore hub to 483\,\tceq. A hub in Amsterdam (Rome) would contribute 27 (53) \tceq, bringing the meeting total to 1,201 (1,228) \tceq. This third meeting hub would bring about a total reduction of $\sim$65\% in emissions. 
\item 4-hub model. A far-Eastern hub in Tokyo (Beijing) would generate 68 (60) \tceq\ from nearby participants (including Australia), and the emissions associated with the Los Angeles hub would drop from 692 to 472 \tceq. The total meeting emissions would be roughly 1,060\,\tceq, resulting in an overall reduction of 70\%. 
\end{itemize}

From this illustration, it is clear that the hub model for conferences can bring significant reductions in emissions due to travel, although we should note that the biannual AAS conference is not a typical astronomy meeting and therefore the extent of the reductions delivered by a hub model may also not be typical. Moreover, the AAS meeting location rotates around the USA every 6 months in order to visit different constituencies, with Seattle being one of those location. Other venues were located more centrally (e.g., Denver, CO), but some very remotely (e.g., Honolulu, HI). 
There are other considerations for the hub model, alongside the obvious increase in expenditure by hosting the meeting in multiple physical locations.
For example, each subsequent hub serves a smaller fraction of attendees and therefore also isolates them increasingly. Because the research fields in astronomy are not defined by location (in contrast to, e.g., geologists, or field biologists who might study a phenomenon in a specific region) the hub concept might not necessarily bring together people who work on similar research topics, and therefore can further enhance the impact of intrinsic geographical disadvantage. 
A rotation of hub locations could support regularly mixing of the pool of attendees such that networking does not become stale.
For national meetings, part of the draw might be to bring an entire country's researchers together at once, in which case a hub model might not be appropriate, but one could look at other strategies for emissions reduction (e.g., meeting every two years instead of annually).
The illustration shows, however, that in certain circumstances even a single additional meeting location (the 2-hub model) can bring about a substantial reduction in emissions, and it might be suitable to consider for large meetings, e.g., the AAS meetings, which are not highly specialised conferences, but attract a big crowd of researchers, such that enough opportunities for networking are given. 

The adoption of a hybrid in-person/online model can bring further carbon reductions.
Previous studies\citep{Bousema2020,Periyasamy2022,burtscher2020,kloewer2020,moss2021_na,Tao2021, Gratiot2023} have shown that hybrid meetings drastically reduce emissions and virtual meetings reduce the amount of emissions by more than 99\% compared to in-person meetings. Three AAS meetings in 2020--21 were held virtually out of necessity (COVID-19 pandemic), so it is certainly feasible for the AAS to adopt the virtual model. The January 2022 meeting was due to be held physically in Salt Lake City, UT with limited online presence, but was cancelled at short notice. 
The June 2022 meeting took place as an in-person meeting in Pasadena, CA, but the January 2023 meeting in Seattle was offered as a hybrid meeting.
If we consider the Seattle meeting from January 2019 as an analogue to the January 2023 meeting, and assume that those based in the contiguous USA, Alaska and Canada had attended in person while the remainder (including Hawai'i) joined the meeting virtually, the emissions total comes to 2,881\,\tceq, which corresponds to a saving of 581\,\tceq\ ($\sim$17\%).
This reduction is a somewhat sizeable amount illustrating an initial potential for organisers to decrease astronomy's impact on the climate by adapting hybrid concepts. However, in order to reach the goal of net zero emissions by 2045 and assuming a linear trend of reductions, decarbonization by at $\geq$5\% per year is necessary, which means that a hybrid model can only provide the needed decrease in GHG emissions for roughly three years, unless the fraction of virtual participation continues to increase. Still, in order to adapt and transform our way of conferencing, hybrid meetings might be a valuable step on the way, if they are implemented successfully such that researchers experience them as good choices for meeting participation. Currently, hybrid participation enables the attendance of (keynote) talks and digital access to posters, and networking among virtual and in-person participants is rarely supported, which disincentivizes people from attending a meeting remotely \citep{Moss2022}. By making better hybrid options available in the future, online attendees will become less disadvantaged, and more people would likely elect to attend online, thus leading to more significant reductions of emissions via hybrid formats. 

\subsection{Environment \& Equity}

While travel has been a regular expected activity for many astronomers in recent decades,
it is important to recognise that the requirement to travel for research creates a barrier to participation for many, which disproportionately affects scientists belonging to minority groups \citep{sarabipour20,skiles21,kohler22}.
Funding is harder to obtain for those based in the less wealthy and less-renowned institutes; time is hard to acquire for people with caring responsibilities (predominantly women); and visas can be a hurdle financially, temporally, and politically, depending on one's citizenship.
And as many in the literature have discussed, virtual events can be just as successful as in-person ones through the appropriate use of modern and constantly improving technology and formats
\citep[e.g.,][]{moss2021_na, chloe, zotova, sarabipour, Gunther2021, Anderson2021,Roos2020, Burtscher2021_messenger}.
Here, we want to elaborate briefly on access barriers created by in-person meetings.

A major factor that decides participation at an event is the financial cost. About 70\% of all meetings take place in Europe and North America. About 86\% of the North American meetings take place in the USA. While the costs for a meeting depend on location, venue, and meeting type, prices across Western Europe, the US, and Canada are approximately the same. Meeting organisers charge a broad range of registration fees, from {$\sim$200--950\,\euro} ($\sim$220--1050\,USD\footnote{Assuming 1\,\euro\,=\,1.1\,USD}), with the mid-to-low range being more common.
Hotel prices also vary depending on location, ranging from 50 to {200\,\euro} (55 to 220\,USD) per night.
International flights between Europe and North 
America range roughly from 700\,\euro\ to 1,000\,\euro\ (770 to 1,100\,USD)
for a return ticket, while the price of travelling from or to other parts of the world is generally considerably more expensive. 
Domestic return flights in the US\footnote{Annual U.S. Domestic Average Itinerary Fare in Current and Constant Dollars - Bureau of Transportation Statistics:  \url{https://www.bts.gov/content/annual-us-domestic-average-itinerary-fare-current-and-constant-dollars}} and airfares in Europe\footnote{Average passenger fare of selected airlines in Europe in 2021 -- Statista: \url{https://www.statista.com/statistics/1125265/average-ticket-price-selected-airlines-europe/}} cost $\sim$200-350\,\euro\ ($\sim$220-380\,USD) on average in 2021/2022. 
While in the US, train travel is rarely feasible mainstream means of transport, it is more common in Europe, and a few connections exist that are reasonable replacements for flights (e.g., London--Paris, Paris--Amsterdam, Berlin--Warsaw, Munich--Budapest, Zurich--Milan\footnotemark).
However, train ticket prices are often more expensive than airfare. For example, the airfare of a 500\,km journey within Italy is about a third cheaper than a train ticket \citep{BERGANTINO2020100823}, while more broadly within Europe the same distance is $\sim$2.5 times cheaper. For journeys in Europe exceeding 1,000\,km the ratio of train ticket prices to plane tickets is $\sim$1.5\footnotemark[\value{footnote}].
\footnotetext{Data visualisations by Deutsche Welle. What's the real cost of travel? - Tom Wills: \url{https://github.com/dw-data/travel-cost}}
Altogether, the price for a five-day meeting in the Eurocentric/North American context, without meal costs, can range from $\sim$500\,\euro\ (550\,USD) to over $\sim$\,3000\,\euro\ ($\sim$\,3300\,USD), depending travel distance and price of conference fee and accommodation. For other regions requiring longer-distance travel, these estimates are likely to be lower limits.

\begin{figure}
    \centering
    \includegraphics[width=.48\textwidth]{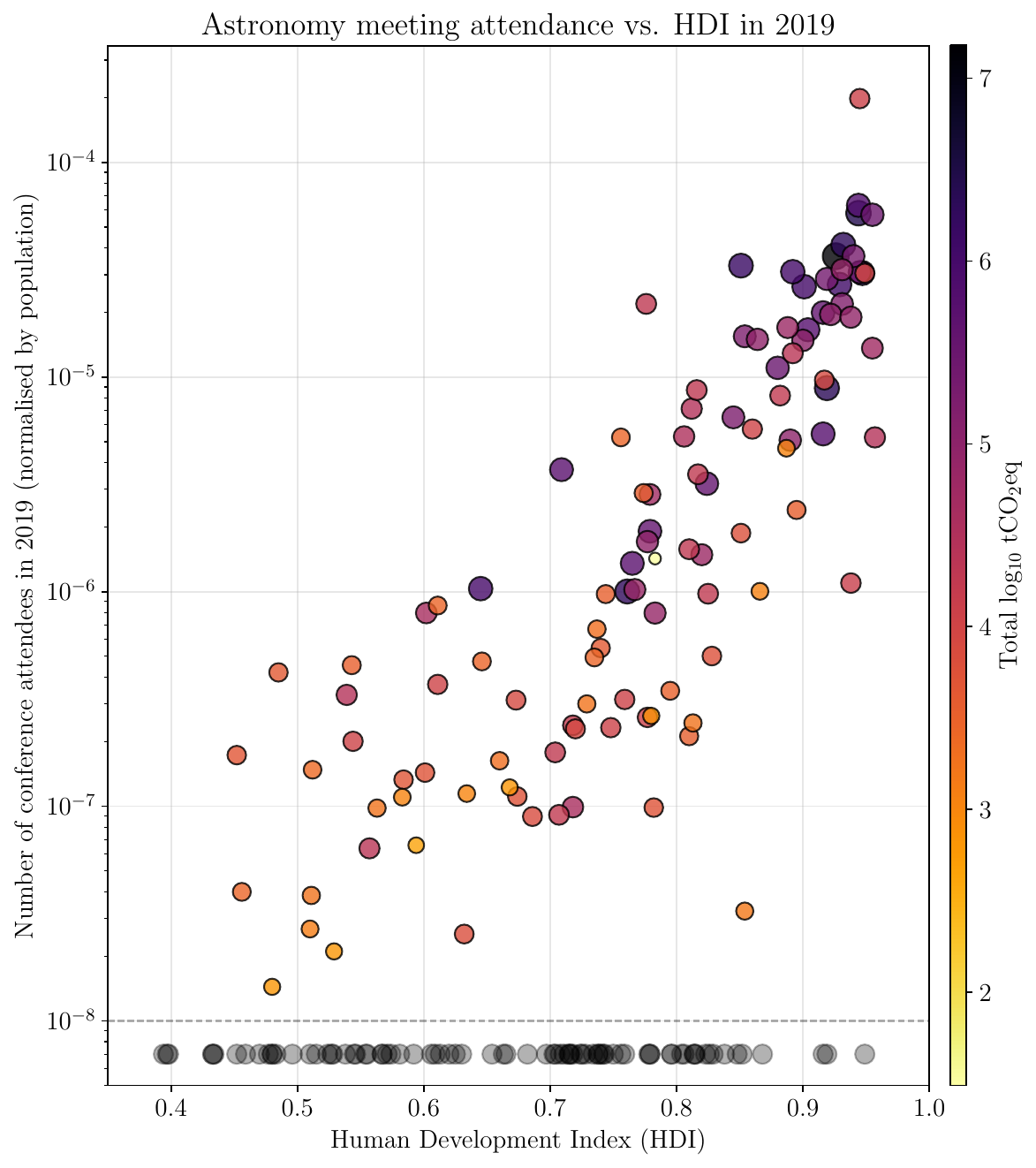}
    \caption{Comparison of the Human Development Index (HDI) of each country of origin for conference participants with the total conference attendees from that country, normalised by population. The color scale and size of the markers show the total amount of GHG travel emissions of all participants for each country. Circles below the dashed line represent the distribution of the HDI for countries with no attendance at astronomy/astrophysics meetings in 2019.}
    \label{fig:human_development_index}
\end{figure}

Hence, in-person participation is largely dependent on available funding, while the distribution of money for work-related travel varies significantly by country, between institutions, and often even within institutes, where senior researchers are known to travel significantly more than early-career researchers \citep{stevens2020}. Certain research facilities and regions are strongly disadvantaged when attempting to take part in in-person scientific meetings simply due to the high cost of travel. Since a large number of 2019 astronomy meetings took place in Europe or North America, researchers in these regions have, in general, the advantage to attend in-person meetings at a lower cost. The privilege of being a researcher in a more advantaged country can also be seen clearly when correlating meeting attendance with Human Development Index (HDI), which seeks to measure human development based on metrics of health, knowledge and standard of living\footnote{UNDP Human Development Reports - Human Development Index: \url{https://hdr.undp.org/data-center/human-development-index\#/indicies/HDI}}. 
Fig.~\ref{fig:human_development_index} shows a scatter plot of the number of 2019 astronomy conference attendees from a given country (normalised by population) as a function of HDI. There, the total carbon emissions for each country, created by travelling to conferences, is expressed by the colourscale and the size of the marker, while countries which did not have any attendees in our dataset are shown at the bottom as circles beneath the dashed line. A clear correlation is seen between conference participation and the development state of a country, such that countries with higher HDI have many more attendees at in-person astronomy conferences. At the same time, researchers in these same high-HDI countries produce more GHG emissions through travel, and contribute proportionally more to the carbon footprint of astronomy. It is also worth noting that 81 countries were missing from the 2019 conference attendee population, meaning that many countries are excluded entirely from in-person conference participation. The average HDI of countries which did feature in our dataset is 0.77$\pm$0.14, while the average HDI of those which did not is 0.66$\pm$0.14, which supports the notion that countries with less development (and thus, less privilege) are consequently less able to send participants to in-person meetings. 

%

Another access barrier to in-person meetings --- which can be invisible to researchers with travel-privileged nationalities --- is the need for long and complicated visa processes. Researchers from Asia and Africa encounter visa-related obstacles for work-related travel far more often than their colleagues from North America or Europe \citep{waruru2018}. Political decisions such as travel bans for citizens of specific countries and month-long waiting times for visa interviews can actively exclude researchers from attending in-person meetings, as well as disadvantage them \citep{ebrahimi2022}.
At least, this is true for a model where in-person meetings are the only option for participating in essential career-advancing academic networking \citep[e.g.,][]{wang2017,campos2018,heffernan2021}.
With the shift to virtual meetings during the COVID-19 pandemic, data could be gathered on how the attendance at international meetings changed with regard to diversity. While, in general, a significant increase in participant numbers occurred \citep[e.g.,][]{stevens2023}, researchers from many more countries, particularly in the global South, registered for the virtual meetings compared to the same major conferences that were held prior to the pandemic \citep{wu2022}. In addition, the participation of some underrepresented minorities (Hispanics and people with African heritage) slightly increased.

The struggle to balance work and care-taking responsibilities is not unique in academia, but a well-known problem for any family. The climb up the academic career ladder is a competitive undertaking, and experience in an international environment is one of the key aspects that is highly valued by universities \citep{lim2017,herschberg2018}.
Both male and female researchers are affected by the complex problem of combining a career in science with a family life, resulting in having fewer children than desired \citep{ecklund2011}.
In general, the mobility of researchers is lowest at the peak of childbearing age, independent of gender \citep{Viry2015}.
However, mobility to (international) conferences that take place in person seems to be linked to gender, generally disfavouring women, which can in parts be attributed to caring responsibilities \citep{leeman2010,joens2011,henderson2020, henderson2021}.
Efforts exist to counter-balance this effect in order to facilitate participation at meetings for people with child-caring responsibilities \citep{bos2019}, however, a favourable family-friendly option can be to attend meetings virtually, such that time away from home is minimized. This will be particularly true if meetings that take place in online or hybrid formats maintain their full effectiveness compared with traditional in-person formats, and as such emphasises the criticality of avoiding those who attend meetings online being disadvantaged in terms of the benefits offered.

To attend an in-person meeting, one often requires a certain level of physical ability to be able to travel to and partake in a conference. Reference \citep{picker2020} describes in detail the obstacles that a researcher with a physical disability must overcome. The accessibility of a meeting is often not considered during planning, and even if organisers take into account how to make participation easier for disabled researchers, the often much higher attendance costs for them --- e.g., due to special transport, personal assistance by an accompanying person --- require additional financial support.
In the literature, several suggestions and guidelines exist on how to organise in-person meetings that are more accessible \citep{picker2020,callus2017,irish2020,joo2022}. Virtual meetings have the advantages of a great reduction in costs and a schedule that does not demand time or energy in exhausting travel. It is especially important when it comes to accessibility at conferences that those who are impacted are consulted and included in the designing and enacting of suitable solutions.

When considering the cumulative impacts of accessibility and inclusivity in academic meetings, it is worth noting that true systematic change in terms of increased diversity at meetings --- even when shifting to more accessible formats such as online and hybrid --- is unlikely to happen in the short period of a few years. Instead, it will require the commitment of a community in the long term to positive change, via actively removing (and keeping removed) barriers of access.

\subsection{A path forward to more sustainable and inclusive meetings}

Ideally, the reduction of GHG emissions from travel to scientific meetings is conducted such that the quality of research is not impacted.
Meetings are beneficial for collaboration and travel to in-person meetings has been found to increase visibility \citep{Berne2022}, however, it has only a limited effect on success \citep{Wynes2019}, 
and it can also reduce an individual's productivity \citep{Seuront2021}.
A straightforward way of cutting emissions therefore could just be reducing the number of meetings, e.g., organising recurring annual meetings only every other year.
This could also have other benefits such as more available time for research, teaching, or mentoring. In addition, it would impose less pressure to be away from home for the sake of visibility for researchers with families, which is a phenomenon that female scientists encounter more often than their male colleagues\citep[e.g.,][]{henderson2021}.

As humans, we find it easiest to network in person, but by solely relying on in-person meetings, certain groups of people can be excluded from the global science community.
In order to allow for both an in-person experience as well as accessibility, an approach could be a hybrid format in which the visibility and networking opportunities for online attendants are prioritised.
For large meetings, a hub format including virtual participation could be considered. 
While it adds an additional effort on organising committees to establish well-working concepts, we recommend avoiding a 'reinventation of the wheel' and instead taking advantage of already existing resources \citep[see, e.g.,][]{Leochico2021}.
In addition, it is worth changing the concept of a conference when we transition into a virtual meeting space, especially for a global event with participants from different time zones. Even though online meetings are capable of reducing the barriers we mentioned earlier, they are still perceived as less effective for networking and social opportunities \citep{Doshi2014}. Hence, to foster more collaboration and enable new connections, the emphasis should lie on active rather than passive participation, as well as asynchronous schedules to accomodate for different time zones \citep[e.g.,][]{neubauer2021,Counsell2020,Lisiecki2021,Wagner2023}. An example for a successful recurring online meeting is `Cosmology from Home', which has been implemented as a virtual conference in 2020 and never went back to the in-person format.
In the future, the advance of technology could even enable hosting both conferences and schools in virtual reality \citep{Wu2019, Taylor2020, moss2021_na}. However, additional skills are necessary to handle the associated hard- and software, compared to simply attending an online meeting via online video conferencing tools, e.g., Zoom or Google Hangouts or Meet.
`Astronomers for Planet Earth' has published `A statement on Conferences and Meetings'\footnote{\url{https://astronomersforplanet.earth/wp-content/uploads/2023/12/A4E-Statement-on-Events-and-Meetings.pdf}} that establishes criteria that should be taken into account for the organization of meetings in order to steer towards a sustainable future.

\section{Conclusions}
The COVID-19 pandemic forced the research community to abruptly change their habit of in-person meetings into virtual ones. 
While this rapid transition has not given us enough time to perfect online meetings, there are many lessons we can learn from that time, and we as the science community should consider planning future meetings with a much larger concern for sustainability, accessibility, and inclusivity. 
Our analysis of in-person astronomy meetings in 2019 finds that the total amount of GHG emission related to travel to open astronomy and astrophysics meetings is \totEmissionsRounded. The average carbon footprint for travel to a conference is \emissionsPerCapConf\ per attendee, and \emissionsPerCapSchool\ per person for travel to a school.
Since academic staff is willing to reduce GHG emissions, it is necessary that universities and research institutions incentivize the shift towards more sustainable practices with regard to long-distance travel, e.g., by removing barriers that currently exist for rail travel.
Currently, no net-zero aviation technology is available, and it will not be possible anytime soon. The amount of Sustainable Aviation Fuels (SAFs) comprised less than 0.1\% of the total fuel consumed in 2019, with the production costs currently being the limiting factor for wider usage \citep{overton2021}. 
Hence, reducing the amount of flights is the most obvious solution to reducing travel-related GHG emissions for the foreseeable future. We address this appeal particularly to senior researchers who, on average, have been found to form the most frequently travelling group \citep{Blanchard2022,stevens2020,ciers2019}.

While purely online meetings have the advantage of drastically decreasing the carbon footprint of an event, we acknowledge that despite newly arising formats this might not be the most suitable option for all meetings. 
Other approaches can include: reducing travel by hosting a hub-conference, hybrid meetings where the needs of online participants are put first, simply optimising the meeting location based on the origin of its participants, or a combination of these. We recognise the logistical challenges that appear for such alternative formats, such as dealing with multiple time zones and the need for more local organisers, but we deem them worthwhile impactful endeavours in the face of the intensifying climate crisis. 
In addition to making the field of astronomy --- and others --- more sustainable, these efforts will also create a more accessible and inclusive research community.

\section{Materials and Methods}\label{sec:matmet}
\subsection{Experimental design}
The goal of this study was to gather data that permit both a qualitative and quantitative analysis of astronomy's conferencing-related travel climate impact. This required a set of astronomy meetings taking place in 2019 that is as complete as possible, as well as inferred travel data for each meeting, such as a list of conference participants and their origin of travel. For the origin of travel, we assume that travellers conduct a return journey from their home institute (i.e., their primary affiliation at the time). We acknowledge that some medium- to long-distance travellers might have combined several conferences or institute visits into one long-distance trip, but since we do not track individuals, we do not account for this in our calculations.

\subsection{Meeting sources and collected data}
As stated above, the target meetings are \textquoteleft open call\textquoteright~astronomy meetings, that is, meetings that are not restricted to the membership of a particular project or collaboration and have thus been publicly advertised. We define `astronomy meetings' as meetings primarily involving professional astronomers and students in the area of studies relating to `photon-based astronomy'. As a consequence, we excluded meetings predominantly geared towards Solar System planetary science, pure astroparticle physics, theoretical physics, and meetings with a wide topical coverage, where core astronomy or astrophysics topics covered only a small fraction of the meeting. The boundaries for including or excluding certain meetings are somewhat open to interpretation, but the above description of inclusion criteria defines our selection function well enough to be potentially re-applied to similar assessments in future years.

The most prominent collection of international astronomy meeting announcements is maintained by the {\em Canadian Astronomy Data Centre (CADC)}{\footnote{Canadian Astronomy Data Centre -- International Astronomy Meetings: \url{https://www.cadc-ccda.hia-iha.nrc-cnrc.gc.ca/en/meetings/}}}, from which we selected meetings that were held in the 2019 calendar year. After an online search for other resources, we included further meetings from {\em Exoplanet.eu}{\footnote{Exoplanet.eu - Meetings: \url{https://exoplanet.eu/meetings/past/}}}, 
the two AAS annual meetings{\footnote{AAS - Meeting Archive: \url{https://aas.org/meetings/past-meetings}}}, the EAS annual meeting, and collections of meetings from other communities e.g.\ 
the {\em Indian Institute of Astrophysics}{\footnote{IIA - Events: \url{https://www.iiap.res.in/?q=events/meetings}}} and the {\em Astronomical Society of Australia}{\footnote{ASA - Events Calendar: \url{https://asa.astronomy.org.au/events/calendar}}}.
After applying the stated selection criteria and removal of two predatory conferences, as well as a few that had been cancelled, this resulted in a total of \totMeetings\ meetings.
To put the number of meetings for 2019 in context to the number of meetings in previous years, we compare those numbers for meetings announced on the websites of the CADC and {\em Exoplanet.eu}. We find that in 2019 the number of meetings is on average $\sim$20\% higher than for those announced in the years 2011 through 2018. Partially, this is also due to the significant increase in exoplanet meetings, for which $\sim$53 were announced on average from 2011 to 2018, but 84 in 2019.

For each of these meetings we searched the meeting websites for lists of participants, which record the number of attendees, as well as the city and country of the primary affiliation of each attendee, i.e., the city we assumed as the travel origin. Where participant lists were not publicly available, we contacted the organisers to ask for an anonymised list, which disclosed at most the affiliation for each participant, or 
at least the state or country for said affiliation. 
We obtained data for participants origins for \totGoodConf~conferences and \totGoodSchool~schools.
Where we did not have access to such a list and did not manage to get in contact with the organisers, we attempted to estimate an approximate number of participants from e.g.\ conference reports or conference photos. For these cases we did not record cities of travel origin, but instead multiplied the number of participants by the continent-averaged per-capita emission value from Table~\ref{tab:general_statistics} to get the footprint of the meeting --- see Table~\ref{tab:coldcases}. For cases where we could not even retrieve the number of participants, we assigned these meetings the continent-averaged number of participants, and multiplied this number by the continent-averaged per-capita emission value from Table~\ref{tab:general_statistics} to get an estimate for the footprint of the meeting. We list the \totColdCaseConf~conferences and \totColdCaseSchool~schools in Supplementary Table 3 \& 4 and discuss the impact of these assumptions below.

We would like to stress that we explicitly did not trace people or institutions in this process. Any participant data was immediately converted to city and country of work and only city and country information were used in any analysis. We did not aggregate information of individual participants or institutions across meetings, nor did we, at any point, associate emission information with individuals. Our interest in this analysis did not lie in individual behaviour, or even the behaviour of specific sub-disciplines of astronomy, but in information on the `meeting sociology' of the joint astronomical community.

\subsection{Conferences and schools}

Astronomy meetings seem to fall into two broad categories, that is conferences and schools. Usually, conferences are aimed at the entire astronomical community, with the primary purpose of presenting and discussing scientific findings in talks or posters. Schools are held for the purpose of training or educating students or early-career scientists in a specific topic or skill. Content is usually delivered in the form of lectures, workshops, hackathons, or other hands-on practices. Our meeting sample consists of \totGoodConf\ conferences with complete participation data, \totColdCaseConf\ conferences without complete data, \totGoodSchool\ schools with complete participation data and \totColdCaseSchool\ without complete data.
We note that we assigned some meetings labelled as `workshop' to the conference category when their program revealed that the format was more conference- than school-like, i.e., mainly consisting of talks presenting results or methods.

\subsection{Calculation of CO$_2$-equivalent emission for journeys}

The major outcome of our collection of participant origin data is an estimate of climate-relevant emissions for all travels to each conference. Since no further data on routes, travel modes, travel companies, travel classes, or airplane models are available, we made some simplifying assumptions. We discuss the potential impact of these in the section \textquoteleft Statistical Analysis\textquoteright~below. 

The tool we used to calculate emissions is the `Travel Carbon Footprint Calculator' by Didier Barret \citep{barret2020}. This online calculator simplifies travel to a direct route between origin and destination, using the geolocation of the two places. It differentiates trips into train and flight trips.
In order to account for deviations from Great Circle distances, a multiplication factor of 1.35 is applied by the tool to all land-bound travel routes.
As a simplification we treat all ground travel as train rides, ignoring the higher emissions of cars. In this way we will slightly underestimate the \COT~emissions.
However, as conference attendance by car involving a drive of more than $\sim$ 300\,km is typically rare and the main emissions stem from medium- to long-distance flights anyway, this will hardly impact the overall picture.

The default maximum distance travelled by train can be set in steps between 0\,km (i.e., not considering train travel at all), 100\,km, 300\,km, 500\,km, 700\,km, and 1,000\,km. We selected the threshold values according to assumed traveller behaviour based on national or continental location. For instance, in mainland Europe, where the rail network is generally good, we generally assumed that for journeys of less than 500\,km, travellers would take the train. Similarly, in North America, where the driving culture is strong and the train network is generally poor, we assumed that travellers would drive journeys of less than 300\,km (but with emissions equivalent to that of trains over the same distance) and aeroplanes otherwise. However, in countries with excellent long-distance rail connections such as China and Japan, we assumed a much larger threshold (700 or 1,000\,km). For meetings held on islands (e.g., Sicily), participants were assumed to fly if they originated from further than the approximate size of the island, even if this distance was within the typical distance threshold for the nation. For example, a trip from Rome to Palermo would be classed as a flight, despite the distance falling under the typical threshold of 500\,km. However, the journey Catania--Palermo ($\sim$200\,km) would be classified as a train journey. 

To get an estimate on how much our choice of train-travelled distance influences the calculations, we calculated CO$_2$-equivalent emissions with both the choice below and the choice above for our picked distances travelled by train, for 36 of the \totGoodMeetings~meetings, e.g., for meetings that we set the maximum distance to 300\,km, we compute also the emissions for 100\,km and 500\,km, etc. We find that the emissions vary, on average, by $1.7$\%. In three out of 71 alternative calculations, the emissions differed between 10 and 20\%. Hence, we conclude that our derived carbon footprint for \totGoodMeetings~astronomy meetings is reasonably well constrained.

The carbon intensity of `train' trips in the tool is assumed to be 23\,g\COT\,km$^{-1}$ per person, which is representative for French trains. `Flight' trips are modelled by a selection of different flight emission models. These differ in the way they include radiative forcings from the non-carbon effects of flights at high altitude, e.g., contrails and cloud creation. The details are described in the calculator paper \citep{barret2020} and on the webpage%
\footnote{\url{https://travel-footprint-calculator.irap.omp.eu}}. 
All three default models that we use in combination (ADEME, myclimate, DEFRA) consider direct emission by aircraft fuel burning as well as upstream emissions for fuel productions\footnote{ADEME: ``Documentation des facteurs d'émissions de la Base Carbone'' via \url{https://base-empreinte.ademe.fr/documentation/base-carbone}; myclimate: \url{https://www.myclimate.org/en/information/about-myclimate/downloads/flight-emission-calculator}; DEFRA: \url{https://www.carbonkit.net/categories/DEFRA_journey_based_flight_methodology}}. 
We note that we used the calculator from July 2021 until December 2022 with the at-that-time emission factors.
Similar to all models implemented in emission calculators, neither model includes the in-comparison negligible emissions from the construction of airplanes \citep{HOWE2013}, nor emissions arising from airport infrastructure, which would add $\sim$4--7\% of emissions over the lifecycle of a plane \citep{jemiolo2015}.
The `RFI' (radiative-forcing index) for the combined three models used is 1.9--2.0, which corresponds to an equivalent to $\sim$165\,g\COT\ per passenger per kilometer for long-distance flights.
There are data supporting higher RFIs: 2.6 or even 3.0 \citep{kloewer2021,lee2021}. For an RFI of 3, our calculated carbon footprint for each meeting would increase by a factor of 1.5. Consequently, our estimates on climate impact are conservatively low -- but could be easily corrected for different model parameters if needed.

For this study, GHG emissions related to train travel are computed based on the typical value for French trains.
However, all countries have a different energy mix, and some might produce electricity with fossil fuels, or even run trains with diesel. In addition, land-bound travel in the US is considered to be done via train while only few regions are well-connected by rail and short distances are therefore travelled by car. 
We can roughly estimate the impact of the simplified approach on the results: 
in total, ground-based travel makes up roughly 3,536,000\,km\footnote{We only take into account data processed with the `Travel Carbon Footprint Calculator' as we have no knowledge about travel routes for meetings for which we lacked participant information.} (taking into account the multiplication factor of 1.35 that accounts for detours of the land routes), which is 1.6\% of the total distance travelled. Applying the emission factor of 23\,g\COT\,km$^{-1}$ per passenger, this yields about 81\,\tceq~for all ground-based travel combined, contributing 0.2\% to the travel emissions of 38,040\,\tceq~(excluding meetings for which we lacked participant information). Using the upper limit for the range of GHG contributions through train travel\footnote{\url{https://www.iea.org/energy-system/transport/rail}}, which is 88.39\,g\COT\,km$^{-1}$ per passenger, the ground-based travel emissions would rise up to 313\,\tceq, which is equivalent to 0.8\% of all emissions, and would add a maximum of 232\,\tceq~to the overall GHG emissions.
Hence, assumptions for train travel are mostly of second-order importance.
We note that other factors, such as the average passenger load as well as the chosen travel class, have an influence on the carbon footprint of each journey, especially for air travel (e.g., flying Business class is associated with higher emissions by a factor of $\sim$\,2 \citep{bofinger2013}). Because we are not able to obtain such detailed information for our study, we assume that all flights were done in Economy class and with 100\% occupancy. Therefore, our results are likely a modest underestimation.

\subsection{Statistical Analysis}

In terms of data collection, there is no complete source of astronomy meetings in a given year. The total number of meetings (\totMeetings) is therefore a lower limit. As noted above, \totColdCaseMeetings\ meetings did not have available participant lists, and of the \totGoodMeetings\ meetings with participant lists, $\sim$10\% were somewhat incomplete (i.e., affiliations were not available or not retrievable for some participants, meaning that cities or countries of origin had to be assumed).
Of course, travel is not the only source of GHG emissions connected with academic meetings in astronomy. Venues themselves often have large carbon footprints; and accommodation is usually required for the meeting participants. We discuss some quantities related to accommodation in the Supplementary Materials, but since they are substantially sub-dominant we do not include accommodation (or venue-specific emissions) in our quantitative estimates presented above.

\subsection{Systematic uncertainties}\label{sec:systematic_unc}

Given the defined reproducible methodology for \COT-emissions, the absolute value of these emissions depends on the chosen RFI value (1.95 in this work) and the simplification of direct routes between travel origin and destination, as well as airtravel classes. The RFI encodes the knowledge uncertainty and model-dependencies of indirect effects of aviation and fuel burning at altitude. The RFI is simply a parameter that could be updated when it is better quantified. Depending on how these emissions numbers shall be further used, either the parameters in \COT-calculation can remain fixed -- maybe for comparisons with subsequent years -- or the RFI and/or direct-flight assumptions can be modified (e.g., for comparison with absolute values of other emission sources). Including a more complex calculation using intermediate stops for long-distance flights instead of great-circle distances will only add a small amount. The Intergovernmental Panel on Climate Change (IPCC) reports a 10\% difference for actual vs.\ great circle-routes \citep{ipcc1999}.

This means that the numbers we are presenting are both a slight underestimate, as well as dependent on the choice of RFI. However, it will depend on the goal for what this number is supposed to be used; whether these dependencies are really a systematic uncertainty or just a result of a specific choice of parameters. For the use as a reference year of conference travel the emissions calculated here are a robust baseline for future comparisons.

\section*{Data availability}
Our calculated data files computed with the 'Travel Carbon Footprint Calculator' and a demonstration of the data analysis are available on Github at \url{https://github.com/agokus/travel-emissions}.

\section*{Supplementary Material}
Supplementary material is available at PNAS Nexus online.

\section*{Acknowledgments}
We want to thank the referees for asking helpful, in-depth questions and making constructive suggestions allowing us to improve the manuscript.
In addition, we would like to thank {\em atmosfair gGmbH} for providing hotel emission data, S.~M.~Wagner for providing helpful comments that improved the manuscript, S.~Falkner for the support in analysing the data with Panda DataFrames, and D.~Barret for background information on the IRAP carbon footprint calculator.
Furthermore, we would like to thank the people who aided in 
data collection and without whom this paper would not have been possible: 
R.~Akeson, I.~Aleman, J.~Alves, E.~Audit, M.~Baaz, J.~Babu,	C.~Bailer-Jones, V.~Beckmann, M.~Bedell,	O.~Beltramo-Martin
% ...

\section*{Author contributions}
    AG and KJ conceptualized the idea behind this investigation and wrote the majority of the manuscript. KJ, AG, PMW, VAM, VOO, ES, ARHS, HD, CK, TAR, JW contributed towards the data collection, sanity checking of the data, analysis and visualisation. LB, VG, and JR 
    were involved in discussions. 

\section*{Funding}
This research did not receive external funding.

\section*{Competing interests}
All authors declare that they have no competing interests.

\bibliographystyle{aasjournal}
\bibliography{flyingreduction}

\end{document}